\documentstyle[preprint,aps]{revtex}
\begin{document}

\draft

\title{Low-temperature behavior of a Magnetic Impurity in a Heisenberg Chain}

\author{Yu-Liang Liu}
\address{Max-Planck-Institut f\"{u}r Physik Komplexer Systeme, Bayreuther
Str. 40, D-01187 Dresden, Germany.}

\maketitle

\begin{abstract}
Using the bosonization technique, we have studied a
spin-1/2 magnetic impurity in Heisenberg chain, and shown that the impurity
specific heat and spin susceptibility have an anomalous temperature
dependence. The temperature dependence of the impurity specific heat is:
 $C_{im}(T)\sim T^{\mu}, \; \mu=4-\frac{1}{g}(1-g)^{2}, \; \text{for} \; 
g_{c}<g< 1$.
The impurity spin susceptibility has the
following temperature dependence: $\chi_{im}(T)\sim T^{\nu}, \;
\nu=3-\frac{1}{g}(1-g)^{2}, \; \text{for} \; g_{c}<g<1$, and
$\nu=-1$, for $g\leq
g_{c}$, where $g_{c}$ satisfies: $4g_{c}=(1-g_{c})^{2}$, and 
$g$ is a dimensionless coupling strength parameter.
\end{abstract}
\vspace{1cm}

\pacs{78.70.Dm, 79.60.Jv, 72.10.Fk}


Recently, 
the quantum impurity scattering of the Tomonaga-Luttinger(TL) liquid has been
extensively studied by using different
techniques~\cite{1,2,3,3d,3',4,5,6,7,8,9,10,11,12,13,13',14,15,16}. 
There is some controversy on the treatment of backward scattering of 
the conduction electrons on a quantum impurity or impurity-like hole in 
the valence band.
One thinks that including backward scattering 
drastically changes the properties of a TL-liquid. 
In principle, this problem can be formulated by means of the Bethe-Ansatz to
obtain exact results for static and thermodynamic quantities.   
However to obtain frequency dependent quantities one has to use 
perturbative methods. 
Due to strong correlations between the conduction electrons and impurity,
perturbation-theory may fail.   
In Ref.~\cite{16}, we have given a clear expression and some exact results
within the validity of the bosonization method for a quantum impurity
scattering of a general one-dimensional(1D) interacting electronic system, and
shown that the system has two independent collective modes: one is drastically
influenced by electron-electron interaction, and another is independent of the
electron-electron interaction which is clearly observed in the numerical
calculations~\cite{15}, while it is missed in the renormalization group
calculations~\cite{3}. 
Here we use the same method as that in
Ref.~\cite{16} to study a magnetic impurity in the
Heisenberg chain which can be easily treated by numerical methods. 
The impurity susceptibility of this system shows an unusual
temperature dependence: 
$\chi_{im}(T)\sim T^{\nu}, \; 
\nu=3-\frac{1}{g}(1-g)^{2},\; \text{for}\; g_{c}<g< 1$, and
$\nu=-1$, for $g\leq g_{c}$, 
where $g_{c}$ is defined as that: $4g_{c}=(1-g_{c})^{2}$, and 
$g$ is a dimensionless coupling strength parameter. 
For an antiferromagnetic Heisenberg chain, the dimensionless coupling strength
parameter $g$ takes the value~\cite{17} 
$g=\frac{1}{2}$, and
the temperature dependence of the impurity susceptibility $\chi_{im}(T)$ 
is $T^{5/2}$. 
The impurity specific heat $C_{im}(T)$ has the temperature
dependence: 
$C_{im}(T)\sim T^{\mu}, \; \mu=4-\frac{1}{g}(1-g)^{2}, \; \text{for}\; 
g_{c}<g<1$. 
Although the impurity specific heat as well as the spin susceptibility depend
on the dimensionless coupling strength parameter $g$, we can still
define a temperature independent Wilson-ratio in this system.   

We consider the following one-dimensional Heisenberg model
\begin{equation}
H=-\sum_{i}\left[
J_{\bot}(S^{x}_{i}S^{x}_{i+1}+S^{y}_{i}S^{y}_{i+1})+
J_{z}S^{z}_{i}S^{z}_{i+1}
\right]
\label{1}
\end{equation}
where $J_{\bot}$ is the transverse exchange interaction strength, and $J_{z}$
the longitudinal exchange interaction strength. 
Using the Jordan-Wigner transformation to spinless fermion operators $f_{i}$,
$S^{+}_{i}=S^{x}_{i}+iS^{y}_{i}=f^{+}_{i} \exp\left[ 
-i\pi \sum_{l<i}n^{f}_{l} \right] , \;\;
S^{z}_{i}=f^{+}_{i}f_{i}-\frac{1}{2}$
and after a Fourier-transformation, the Hamiltonian can be written in the form
\begin{equation}
H=-J_{\bot}\sum_{k}\cos(ak)f^{+}_{k}f_{k}-\frac{J_{z}}{N}\sum_{k}\cos(ak)
\rho(k)\rho(-k)
\label{2}
\end{equation}
where $a$ is the lattice constant, $N$ is the site number, 
the sum over $k$ is restricted to the first
Brillouin zone, $f_{k}=\frac{1}{\sqrt{N}}\sum_{j}e^{ikx_{j}}f_{j}$, and
$\rho(k)$ is the density operator,
$\rho(k)=\sum_{j}e^{ikx_{j}}f^{+}_{j}f_{j}$. 
One expects the
asymptotic behavior of correlation functions to be determined by the low-lying
excited states near the Fermi points at $ak_{F}=\pm\frac{\pi}{2}$. 
Therefore one can consider a related model with two linear single-particle
spectra tangent to the $-\cos(ak)$ at the Fermi points. We may introduce
$f_{1k}(f_{2k})$ operators to describe the fermion particles with positive
(negative) group velocity and the associated fields,
$\psi_{i}(x)=\frac{1}{\sqrt{L}}\sum_{k}f_{ik}e^{ikx},\;\; i=1,2$, where $L=aN$
is the length of the system. The density operator is 
$\rho(k)=\rho_{1}(k)+\rho_{2}(k)$ 
with $\rho_{i}(k)=\sum_{p}f^{+}_{ik+p}f_{ip}$
for $i=1,2$.
For the case of spinless fermions, the $J_{z}$-term in Eq.~(\ref{2}) is
$
\sum_{k\sim 2k_{F}}\rho(k)\rho(-k)\rightarrow -\sum_{k\sim
0}\rho_{1}(k)\rho_{2}(-k) .
$
Therefore the Hamiltonian (\ref{2}) can be simplified as
\begin{equation}
H=v_{F}\sum_{k}k(f^{+}_{1k}f_{1k}-f^{+}_{2k}f_{2k})-\frac{4J_{z}}{N}
\sum_{k}\rho_{1}(k)\rho_{2}(-k) \; . 
\label{3}\end{equation}
The Jordan-Wigner transformation for the spin operators on the lattice has the
obvious generalization to the continuum situation~\cite{17}
\begin{eqnarray}
f_{i} 
&\rightarrow &
(\frac{a}{2})^{1/2}[\psi_{1}(x)+\psi_{2}(x)]  \label{4} \\
i\pi\sum_{l<i}f^{+}_{l}f_{l}
&\rightarrow &
i\pi
\hspace{-0.5em}\int\limits^{x_{i}-a}_{-\infty} \hspace{-0.5em}
dy[\rho_{1}(y)+\rho_{2}(y)+
(2a)^{-1}]\equiv iN(x_{i})	\nonumber
\end{eqnarray}
and the resulting representation for continuum spin operators is
\begin{eqnarray}
S^{-}(x)&=& (\frac{1}{2a})^{1/2}[\psi_{1}(x)+\psi_{2}(x)]e^{iN(x)} \nonumber\\
S^{+}(x)&=& [S^{-}(x)]^{+}  \label{5}	\\
2S^{z}(x)&=&
\rho_{1}(x)+\rho_{2}(x)+\psi^{+}_{1}(x)\psi_{2}(x)+\psi^{+}_{2}(x)\psi_{1}(x) 
\nonumber
\end{eqnarray}
These representations of the spin operators are different from that used in 
Ref.~\cite{b17}. 
Different results of Ref.~\cite{b17} may come from the special
choice for the spin operators and the Kondo interaction between the impurity
and the conduction electrons.
It is well-known that the bosonization representations of the fermion fields
are~\cite{17,18,19}
\begin{eqnarray} 
\psi_{i}(x) &= & \frac{e^{\pm
ik_{F}x}}{\sqrt{2\pi\alpha}}e^{i\Phi_{i}(x)}	\label{6}	\\
\Phi_{i}(x) &= & \mp i\frac{2\pi}{L}\sum_{k}
\frac{e^{-\frac{\alpha}{2}|k|-ikx}}{k}\rho_{i}(k) \nonumber
\end{eqnarray}
where the negative sign for $i=1$ and the positive sign for $i=2$.
$\alpha^{-1}$ is an ultraviolet cutoff, which is of the order of the
conduction band width.
The Hamiltonian (\ref{3}) can be written in terms of boson density
operators 
\begin{eqnarray}
H &=& \frac{2\pi v_{F}}{L}\sum_{k>0}[\rho_{1}(-k)\rho_{1}(k)+
\rho_{2}(-k)\rho_{2}(k)]	\nonumber \\
 && -\frac{4J_{z}}{L}\sum_{k}\rho_{1}(k)\rho_{2}(-k) \nonumber\\
 &=& \frac{v_{F}}{8\pi}(1-\gamma^{2})^{\frac{1}{2}}\int dx[(\partial_{x}
\tilde{\Phi}_{+}(x))^{2}+(\partial_{x}\tilde{\Phi}_{-}(x))^{2}]
\label{7} 
\end{eqnarray}
where
$[\rho_{1}(-k), \rho_{1}(k')] = [\rho_{2}(k'), \rho_{2}(-k)]
=\frac{kL}{2\pi}\delta_{kk'}$; $\Phi_{\pm}(x)=\Phi_{1}(x)\pm\Phi_{2}(x)$, 
$\tilde{\Phi}_{+}(x)=(\frac{1+\gamma}{
1-\gamma})^{1/4}\Phi_{+}(x)$, $\tilde{\Phi}_{-}(x)=(\frac{1-\gamma}{1+\gamma}
)^{1/4}\Phi_{-}(x)$, $\partial_{x}\Phi_{\pm}(x)=2\pi (\rho_{1}(x)\pm
\rho_{2}(x))$,
$\gamma=-2J_{z}/(\pi v_{F})$.
The boson fields $\tilde{\Phi}_{\pm}$ are conjugate variables, $[\partial_{x}
\tilde{\Phi}_{+}$, $ \tilde{\Phi}_{-}(y)]=i4\pi\delta(x-y)$, $
[\partial_{x}\tilde{\Phi}_{-}(x)$, $ \tilde{\Phi}_{+}(y)]=i4\pi\delta(x-y)$.

We consider a spin-1/2 magnetic impurity scattering on the conduction
fermions
\begin{equation}
H_{im}=\tilde{J}^{\bot}_{0}(S^{+}S^{-}(0)+S^{-}S^{+}(0))+\tilde{J}^{z}_{0}
S^{z}S^{z}(0)
\label{8}\end{equation}
By taking $S^{+}=\tilde{f}^{+}, \;\; S^{-}=\tilde{f}, \;\;
S^{z}=\tilde{f}^{+}\tilde{f}-\frac{1}{2}$, the Hamiltonian (\ref{8}) can
be written as
\begin{eqnarray}
\lefteqn{
H_{im} =
J^{\bot}_{0}[\tilde{f}^{+}(\psi_{1}(0)+\psi_{2}(0))e^{iN(0)}+h.c.]
} \label{9}\\
&& +  J^{z}_{0}(\tilde{f}^{+}\tilde{f}-\frac{1}{2})[\rho_{1}(0)
\! + \! \rho_{2}(0) \! + \! 
\psi^{+}_{1}(0)\psi_{2}(0)+\psi^{+}_{2}(0)\psi_{1}(0)] 
\nonumber 
\end{eqnarray}
where $J^{\bot}_{0}=\tilde{J}^{\bot}_{0} / \sqrt{2a}$ and 
$J^{z}_{0} = \tilde{J}^{z}_{0} / 4 $. 
However, due to the appearance of the cross term $\psi^{+}_{1}(0)\psi_{2}(0)+
\psi^{+}_{2}(0)\psi_{1}(0)$ in (\ref{9}), which describes the backward
scattering of the magnetic impurity on the conduction fermions, 
the problem becomes unsolvable. 
Therefore we have to find a 
method to eliminate this cross term. 
For this purpose,  
we define a new set of fermion operators 
\begin{eqnarray}\label{10}
\Psi_{1}(x)&=&\frac{1}{\sqrt{2}}(\psi_{1}(x)+\psi_{2}(-x)) \\
\Psi_{2}(x)&=&\frac{1}{\sqrt{2}}(\psi_{1}(x)-\psi_{2}(-x))  \nonumber 
\end{eqnarray}
and the density operators 
$
\tilde{\rho}_{1(2)}(x) = \frac{1}{L}\sum_{p}\tilde{\rho}_{1(2)}(p)e^{ipx} 
$
and 
$
\tilde{\rho}_{1(2)}(p) = \sum_{k}\Psi^{+}_{1(2)}(k+p)\Psi_{1(2)}(k)
$
which have the standard commutation relation of the right mover,
$[\tilde{\rho}_{1(2)}(-q),\;\;
\tilde{\rho}_{1(2)}(q')]=\frac{qL}{2\pi}\delta_{qq'},\;\; 
[\tilde{\rho}_{1}(q),\;\; \tilde{\rho}_{2}(q')]=0$. Now the bosonization
representation of the fermion fields $\Psi_{1(2)}(x)$ can be performed in the 
standard way~\cite{17,18,19}
\begin{eqnarray}\label{11}
\Psi_{1(2)}(x) &= &
\displaystyle{\frac{1}{\sqrt{2\pi\alpha}}e^{-i\phi_{1(2)}(x)}} \\
\phi_{1(2)}(x) &= & \displaystyle{i\frac{2\pi}{L}\sum_{p}
\frac{e^{-\frac{\alpha}{2}|p|-ipx}}{p}\tilde{\rho}_{1(2)}(p)} \nonumber
\end{eqnarray}
where $\partial_{x}\phi_{1(2)}(x)=2\pi\tilde{\rho}_{1(2)}(x)$. Using
Eq.~(\ref{10}) to express the density operators $\rho_{1(2)}(x)$
\begin{eqnarray}
\rho_{1}(x) &= & \frac{1}{2}[\tilde{\rho}_{1}(x)+\tilde{\rho}_{2}(x)+
\Psi^{+}_{1}(x)\Psi_{2}(x)+\Psi^{+}_{2}(x)\Psi_{1}(x)] \nonumber \\
\rho_{2}(x) &= & \frac{1}{2}[\tilde{\rho}_{1}(-x)+\tilde{\rho}_{2}(-x)-
\Psi^{+}_{1}(-x)\Psi_{2}(-x) \nonumber \\
&&	-\Psi^{+}_{2}(-x)\Psi_{1}(-x)] \nonumber \\
\lefteqn{
\psi^{+}_{1}(0)\psi_{2}(0)+  \psi^{+}_{2}(0)\psi_{1}(0) =
\tilde{\rho}_{1}(0)-\tilde{\rho}_{2}(0)	
}		
\end{eqnarray}
and defining new boson fields: 
$\phi_{\pm}(x)=\phi_{1}(x)\pm\phi_{2}(x)$, which
satisfy the commutation relations:
$[\phi^{'}_{\pm}(x),\;\;\phi_{\pm}(y)]=i4\pi\delta(x-y), \;\;
[\phi^{'}_{+}(x),\;\; \phi_{-}(y)]=0$, 
the Hamiltonians (\ref{7}) and (\ref{9}) can be rewritten as
\begin{eqnarray}
H &= & \displaystyle{\frac{v_{F}}{8\pi}\int 
dx\{(\phi^{'}_{+}(x))^{2}+\gamma\phi^{'}_{+}(x)\phi^{'}_{+}(-x)} \nonumber\\
&& +  \displaystyle{(\phi^{'}_{-}(x))^{2}-\frac{4\gamma}{\alpha^{2}}
\cos(\phi_{-}(x))\cos(\phi_{-}(-x))\}}
\label{12}
\end{eqnarray}
\begin{eqnarray}
H_{im} &= & \displaystyle{
\sqrt{2}J^{\bot}_{0}(\tilde{f}^{+}\Psi_{1}(0)e^{iN(0)}+h.c.)}\nonumber\\
&& +  \displaystyle{
\frac{J^{z}_{0}}{2\pi}(\tilde{f}^{+}\tilde{f}-\frac{1}{2})(\phi^{'}_{+}(0)+
\phi^{'}_{-}(0))}
\label{13}
\end{eqnarray}
where 
$\phi^{'}_{\pm}(x)\equiv\partial_{x}\phi_{\pm}(x)$.  
The cross term in Eq.~(\ref{13}) has vanished, but the boson field
$\phi_{-}(x)$  in~(\ref{12}) becomes strongly self-interacting. 
The Hamiltonian~(\ref{13}) can be rewritten as
\begin{eqnarray}\label{14}
H_{im} &= & \sqrt{2}J^{\bot}_{0}(\tilde{f}^{+}\Psi_{1}(0)e^{iN(0)}+h.c.) \\
&& 	+ \frac{J^{z}_{0}(k=0)}{2\pi}
	(\tilde{f}^{+}\tilde{f}-\frac{1}{2})\phi^{'}_{+}(0) \nonumber \\
&&	+\frac{J^{z}_{0}(k=2k_{F})}{2\pi}(\tilde{f}^{+}\tilde{f}-\frac{1}{2})
	\phi^{'}_{-}(0)	\nonumber
\end{eqnarray}
where for simplicity, we use $J^{z}_{0}(k=0)$ and $J^{z}_{0}(k=2k_{F})$ 
to indicate the forward and backward
scattering interaction strength, respectively, because they show different
effects on the system. 
We can cancel the $J^{z}_{0}$ term in~(\ref{14}) by the 
following unitary transformation
\begin{eqnarray} \label{15} 
U = \exp\{
i(\tilde{f}^{+}\tilde{f} &-& \frac{1}{2})
[\frac{g\delta_{+}}{\pi}\phi_{+}(0) +
\frac{\delta_{-}}{\pi}\phi_{-}(0)]
\} 
\end{eqnarray}
where,
$\delta_{+} = \arctan(
\frac{J^{z}_{0}(k=0)}{2v_{F}\sqrt{1-\gamma^{2}}})$, $
\delta_{-} = \arctan(\frac{J^{z}_{0}(k=2k_{F})}{2v_{F}})$, and 
$g=(\frac{1-\gamma}{1+\gamma})^{1/2}$ is a dimensionless coupling strength
parameter. 

However, the backward scattering potential has a drastical influence on the
fermions $\Psi_{1(2)}(x)$, and induces the strong coupling between the fermion
fields $\Psi_{1(2)}(x)$ and $\tilde{f}$ at the impurity site $x=0$. In the strong
coupling limit induced by the backward scattering potential, i.e., the phase
shift $\delta_{-}$ takes the value: $\delta^{c}_{-}=-\pi/2$, 
and taking the gauge
transformations: $\Psi_{1(2)}(x)=\bar{\Psi}_{1(2)}(x)e^{i\theta_{1(2)}}$,
$\theta_{1}-\theta_{2}=2\delta_{-}(\tilde{f}^{+}
\tilde{f}-1/2)$, we can rewrite the total
Hamiltonian as
\begin{equation}\begin{array}{rl}
\bar{H}_{T}= & U^{+}(H+H_{im})U\\
= &  \displaystyle{
\frac{v_{F}}{8\pi}\int dx\{(\bar{\phi}^{'}_{+}(x))^{2}+
\gamma\bar{\phi}^{'}_{+}(x)
\bar{\phi}^{'}_{+}(-x)}  \\
+&  \displaystyle{
(\bar{\phi}^{'}_{-}(x))^{2}+\frac{4\gamma}{\alpha^{2}}
\cos(\bar{\phi}_{-}(x))\cos(
\bar{\phi}_{-}(-x))\}} \\
+& \displaystyle{
\frac{1}{\sqrt{\pi\alpha}}J^{\bot}_{0}(\tilde{f}^{+}e^{-i(\frac{g\delta_{+}}
{\pi}+\frac{1}{2})\bar{\phi}_{+}(0)}U^{+}e^{iN(0)}U+h.c.)}
\label{16}
\end{array}\end{equation}
where $\bar{\phi}_{\pm}(x)=\bar{\phi}_{1}(x)\pm\bar{\phi}_{2}(x)$,
$\partial_{x}\bar{\phi}_{1(2)}(x)=2\pi\bar{\rho}_{1(2)}(x)$,
$\bar{\rho}_{1(2)}(x)=\bar{\Psi}^{+}_{1(2)}(x)\bar{\Psi}_{1(2)}(x)$.
The last term can be easily obtained by perfoming the unitary transformation
$U^{+}\tilde{f}^{+}U$ which contributes the phase factor
$-ig\delta_{+}/(\pi)\bar{\phi}_{+}(0)-i\delta_{-}/(\pi)\bar{\phi}_{-}(0)$.
In the strong coupling limit ($\delta^{c}_{-}=-\pi/2$), by using Eq.(11), we
can obtain the last term in (\ref{16}). 
If we redefine the following new fields: 
\begin{equation}\begin{array}{rl}
\bar{\psi}_{1}(x)= & \displaystyle{\frac{1}{
\sqrt{2}}(\bar{\Psi}_{1}(x)+\bar{\Psi}_{2}(x))}\\
\bar{\psi}_{2}(x)= & \displaystyle{\frac{1}{
\sqrt{2}}(\bar{\Psi}_{1}(-x)-\bar{\Psi}_{2}(-x))
}\label{17}\end{array}\end{equation}
where the bosonization representation of the fermion fields
$\bar{\psi}_{1(2)}(x)$ is $\bar{\psi}_{1(2)}(x)=(\frac{1}{2\pi\alpha})^{1/2}
e^{-i\bar{\Phi}_{1(2)}(x)}$,  
the total Hamiltonian can be
written as
\begin{equation}\begin{array}{rl}
H^{c}_{T} = &  \displaystyle{
\frac{v_{F}}{4\pi}\int dx\{(\bar{\Phi}^{'}_{1}(x))^{2}+
\gamma\bar{\Phi}^{'}_{1}(x)
\bar{\Phi}^{'}_{1}(-x)} 	\\
+&    \displaystyle{
(\bar{\Phi}^{'}_{2}(x))^{2}
+\gamma\bar{\Phi}^{'}_{2}(x)
\bar{\Phi}^{'}_{2}(-x)} 	\\
+&    \displaystyle{
\sqrt{2}J^{\bot}_{0}[\tilde{f}^{+}\psi(0)+\psi^{+}(0)\tilde{f}]}
\label{20}
\end{array}\end{equation}
where
$\psi(0)=(\frac{1}{2\pi\alpha})^{1/2}\exp\{-i(\frac{1}{2}+
\frac{g\delta_{+}}{\pi})(\bar{\Phi}_{1}(0) +
\bar{\Phi}_{2}(0))\}$. After these transformations, the phase factor $N(0)$
disappears in the total Hamiltonian (\ref{20}). If the
dimensionless coupling strength parameter $g$ takes the values, $g\geq 1$, in
the strong coupling limit induced by the forward scattering potential, the
phase shift $\delta_{+}$ satisfies the relation: $\delta^{c}_{+}=-\pi/(2g)$, 
$\psi(0)$ becomes a constant field, the total Hamiltonian (\ref{20}) is very
similar to that in Ref.\cite{m} derived from the quantum dot. The
$J^{\bot}_{0}$-term opens a gap in the energy spectrum of the fermion
$\tilde{f}$,
therefore, in the low temperature and low energy limit, the Green's function
of the fermion $\tilde{f}$ is an exponential decaying function, the impurity
susceptibility exponentially goes to zero as the temperature going to
zero. However, if the dimensionless coupling strength parameter $g$ is less
than one, $g<1$, the phase shift $\delta_{+}$ only takes the value:
$\delta^{c}_{+}=-\pi/2$, 
the temperature dependence of the impurity susceptibility has
a power-law form (see below). 
Near the strong coupling critical point induced by the forward and backward
scattering potentials, we have the leading 
irrelevant Hamiltonian 
\begin{equation}\label{irrelevant}
\Delta H=\lambda(\tilde{f}^{+}\tilde{f}-\frac{1}{2})\phi^{'}_{+}(0)+
\lambda^{'}(\tilde{f}^{+}\tilde{f}-\frac{1}{2})\phi^{'}_{-}(0)
\label{18}\end{equation}
where $\lambda=-v_{F}(\delta_{+}-\delta^{c}_{+})/\pi$
and
$
\lambda^{'}=-v_{F}(\delta_{-}-\delta^{c}_{-})/\pi$.

According to Eq.~(\ref{10}), we have following relations at the point
$x=0$: 
\begin{eqnarray}
\rho_{1}(0)+\rho_{2}(0) = \tilde{\rho}_{1}(0)+\tilde{\rho}_{2}(0)
&=& \frac{1}{2\pi}\phi^{'}_{+}(0) \nonumber \\
\rho_{1}(0)-\rho_{2}(0) &=& 
\frac{1}{\pi \alpha}\cos(\phi_{-}(0)) \label{18'}\\
\psi^{+}_{1}(0)\psi_{2}(0)+\psi^{+}_{2}(0) \psi_{1}(0) &=&
\frac{1}{2\pi}\phi^{'}_{-}(0) \nonumber
\end{eqnarray}
Therefore, in the srong coupling limit, from (\ref{17}), 
(\ref{20}) and (\ref{18'}), we can easily obtain the following correlation
functions 
\begin{eqnarray}
\displaystyle{
<\phi^{'}_{-}(0,0)\phi^{'}_{-}(0,t)>}\sim  &  \displaystyle{
(\frac{1}{t})^{\frac{2}{g}}} \nonumber\\
\displaystyle{
<e^{i\phi_{+}(0,0)}e^{-i\phi_{+}(0,t)}>}\sim  &  \displaystyle{
(\frac{1}{t})^{\frac{2}{g}}} \label{21}\\
\displaystyle{
<\phi^{'}_{+}(0,0)\phi^{'}_{+}(0,t)>}\sim   &  \displaystyle{
(\frac{1}{t})^{2}}\nonumber
\end{eqnarray}
For an antiferromagnetic Heisenberg chain, the exponent $g$ equals
$1/2$ ($\gamma$ is very large). 

From Eq.(\ref{20})
we can obtain the Green's function of the impurity fermion
\begin{equation}
G_{\tilde{f}}(t)=<\tilde{f}(0)\tilde{f}^{+}(t)>\sim
\left\{\begin{array}{ll} 
0, &  \;\; g\geq 1\\
\displaystyle{
(\frac{1}{t})^{2-\frac{1}{2g}(1-g)^{2}}},  &  \;\; g_{c}<g<1\\
\displaystyle{
e^{-i\epsilon_{f}t}}, & \;\; g\leq g_{c}
\end{array}\right.
\label{22}\end{equation}
where $\epsilon_{f}$ is the level of the impurity fermion $f$, and $g_{c}$ is
defined as that: $4g_{c}=(1-g_{c})^{2}$. The physical interpretation of the
special parameter $g_{c}$ is that at this point the self-energy of the
impurity fermion $\tilde{f}$ induced by the interaction term in (19) has a 
linear frequency dependence.
With the help of Eq.(\ref{21}) we see that in Eq.(\ref{18})
for the antiferromagnetic case
$g<1$, the $\lambda$-term is dominant while for the ferromagnetic case $g>1$,
the $\lambda^{'}$-term is dominant. However, in the case of $g>1$, the Green's
function of the fermion $f$ is exponentially decaying (in Eq.(\ref{22}) we
take it as zero in the long time limit), only the $\lambda$-term be relevant.
By using Eqs.~(\ref{21}), 
and (\ref{22}), the
correlation function of $\Delta H$ (Eq.~(\ref{irrelevant})) can be written as
$<(\tilde{f}^{+}\tilde{f}-1/2)(t)\cdot(\tilde{f}^{+}\tilde{f}-1/2)(0)><
\phi^{'}_{+}(0,t)\phi^{'}_{+}(0,0)>$, if we omit the vacuum fluctuation of the
fermion $\tilde{f}$, it reads
\begin{equation}
<\Delta H(0,0)\Delta H(0,t)>\sim\left\{\begin{array}{ll}
\displaystyle{
\lambda^{2}(\frac{1}{t})^{6-\frac{1}{g}(1-g)^{2}}}, & \;\; g_{c}<g<1\\
\displaystyle{
\lambda^{2}(\frac{1}{t})^{2}}, & \;\; g\leq g_{c}
\end{array}\right.
\label{23}\end{equation}
while the spin susceptibility of the impurity is (omitting the vacuum
fluctuation of the fermion $\tilde{f}$)
\begin{equation}
<S^{z}(0)S^{z}(t)>\sim
(\frac{1}{t})^{4-\frac{1}{g}(1-g)^{2}},   \;\; g_{c}<g<1
\label{24}\end{equation}
 From Eqs. (\ref{23}) and (\ref{24}), using the relation between specific heat
and free energy (Eq.(\ref{23}) gives the impurity free energy), 
we can easily obtain the temperature
dependence of the impurity specific heat $C_{im}(T)$ and spin susceptibility
$\chi_{im}(T)$
\begin{equation}
C_{im}(T)\sim\left\{\begin{array}{ll}
\displaystyle{
\lambda^{2}T^{4-\frac{1}{g}(1-g)^{2}}}, & \;\; g_{c}<g<1\\
constant, & \;\; g\leq g_{c}
\end{array}\right.
\label{25}\end{equation}
\begin{equation}
\chi_{im}(T)\sim\left\{\begin{array}{ll}
\displaystyle{
T^{3-\frac{1}{g}(1-g)^{2}}}, &  \;\; g_{c}<g<1\\
\displaystyle{
T^{-1}}, & \;\; g\leq g_{c}
\end{array}\right.
\label{26}\end{equation}
which show an unusual temperature dependence in the low energy and
low temperature limit. 
We can give a brief explanation about present results. It is worth noted that
because we use an usual Kondo interaction as in [21] which is different from
that in [20], we obtain different results from that in [20]. The symmetry of
the Kondo interaction term significantly influences the low temperature
behavior of the magnetic impurity. In generally, the Heisenberg chain can be
described by an interacting spinless electron system. As the interactions
among the electrons are repulsive, the bound state of the conduction electron
and the impurity fermion is weakened by the repulsive interaction of the
conduction electrons. For an enough strongly repulsive interaction
$g=g_{0}$,
where $g_{0}$ satisfies: $3g_{0}-(1-g_{0})^{2}=0$, the bound
state at the impurity site is broken, and the impurity fermion begins 
to show a free-type behavior.  
For an
antiferromagnetic Heisenberg chain, the dimensionless coupling strength
parameter takes
the value~\cite{17} $g=\frac{1}{2}$, the impurity specific heat $C_{im}(T)$ 
is proportional to $T^{7/2}$, and the temperature dependence of the impurity
spin susceptibility $\chi_{im}(T)$ is $\chi_{im}(T)\sim T^{5/2}$. 
Because in the Jordan-Wigner representation, the Heisenberg chain can
be reduced into a very simple form, these anomalous temperature dependences of
the impurity specific heat and spin susceptibility should be easily observed
in the numerical calculations. Although these temperature dependences
heavily rely upon the dimensionless coupling strength parameter $g$, 
we still have a temperature independent Wilson ratio which exists in 
the high-dimensional Kondo problems.

In summary, by using the bosonization method, we have studied in detail 
the low temperature physical
behavior of the spin-1/2 magnetic impurity in Heisenberg chain, and shown that
the impurity specific heat and spin susceptibility have unusual temperature
dependence behavior in the low energy and low temperature limit.

The author would like to thank Prof. P. Fulde for encouragement and
Dr. K. Fischer for valuable discussions.


\begin{references}

\bibitem{1} T.Ogawa, A.Furusaki, and N.Nagaosa, Phys. Rev. Lett. {\bf 68}, 
3638(1992); A.Furusaki, and N.Nagaosa, Phys. Rev. {\bf B}47, 3827(1993).

\bibitem{2} D.K.K.Lee, and Y.Chen, Phys. Rev. Lett. {\bf 69}, 1399(1992).

\bibitem{3} C.L.Kane, and M.P.A.Fisher, Phys. Rev. Lett. {\bf 68}, 
1220(1992); Phys. Rev. {\bf B}46, 15233(1992).

\bibitem{3d} D.H.Lee, and J.Toner, Phys. Rev. Lett. {\bf 69}, 3378(1992);
A.Furusaki, and N.Nagaosa, {\it ibid}., {\bf 72}, 892(1994).

\bibitem{3'} X.G.Wen, Phys. Rev. {\bf B}41, 12838(1990);
Int. J. Mod. Phys. {\bf B}6, 1711(1992).

\bibitem{4} K.A.Matveev, D.X.Yue, and L.I.Glazman, Phys. Rev. Lett. {\bf 
71}, 3351(1993).

\bibitem{5} A.D.Gogolin, Phys. Rev. Lett. {\bf 71}, 2995(1993).

\bibitem{6} N.V.Prokof'ev, Phys. Rev. {\bf B}49, 2148(1994).

\bibitem{7} C.L.Kane, K.A.Matveev, and L.I.Glazman, Phys. Rev. {\bf B}49, 
2253(1994).

\bibitem{8} I.Affleck, and A.W.W.Ludwig, J. Phys. {\bf A}27, 5375(1994).

\bibitem{9} M.Ogata, and H.Fukuyama, Phys. Rev. Lett. {\bf 73}, 468(1994).

\bibitem{10} P.Fendley, A.W.W.Ludwig, and H.Saleur, Phys. Rev. Lett. 
{\bf 74}, 3005(1995).

\bibitem{11} F.Guinea, G.G\'{o}mez-Santos, M.Sassetti, and M.Ueda, 
Europhys. Lett. {\bf 30}, 561(1995).

\bibitem{12} S.Tarucha, T.Honda, and T.Saku, Solid State Commun., {\bf 
94}, 413{1995}.

\bibitem{13} K.Moon {\it et al}., Phys. Rev. Lett. {\bf 71}, 4381(1993); 
K.Moon, and S.M.Girvin, cond-mat/9511013.

\bibitem{13'} F.P.Milliken, C.P.Umbach, and R.A.Webb, Solid State Commun., 
{\bf 97}, 309(1996).

\bibitem{14} Y.Oreg, and A.M.Finkel'stein, Phys. Rev. {\bf B}53, 10928(1996);
Phys. Rev. Lett. {\bf 76}, 4230(1996).

\bibitem{15} F.Lesage, H.Saleur, and S.Skorik, Phys. Rev. Lett. {\bf 76}, 
3388(1996); G.G\'{o}mez-Santos, {\it ibid}, {\bf 76}, 4223(1996).

\bibitem{16} Y.L.Liu, Quantum impurity scattering of Tomonaga-Luttinger
liquid, preprint.

\bibitem{b17} G.Clarke, T.Giamarchi, and B.I.Shraiman, Phys. Rev. {\bf B}48,
7070(1993).

\bibitem{17} A.Luther, and I.Peschel, Phys. Rev. {\bf B}9, 2911(1974);
{\bf B}12, 3908(1975).

\bibitem{18} J.S\'{o}lyom, Adv. Phys. {\bf 28}, 201(1979).

\bibitem{19} F.D.M.Haldane, J. Phys. {\bf C}14, 2585(1981).

\bibitem{m} K.A.Matveev, Phys. Rev. {\bf B}51, 1743(1995).



\end{references}
\end{document}